\documentclass[conference]{IEEEtran}
\IEEEoverridecommandlockouts
\usepackage{cite}
\usepackage{amsmath,amssymb,amsfonts}
\usepackage{algorithmic}
\usepackage{graphicx}
\usepackage{textcomp}
\usepackage{xcolor}
\usepackage{url}
\usepackage[T1]{fontenc}
\def\BibTeX{{\rm B\kern-.05em{\sc i\kern-.025em b}\kern-.08em
    T\kern-.1667em\lower.7ex\hbox{E}\kern-.125emX}}
    \newtheorem{remark}{Remark}

\newcommand{\dif}{\mathop{}\!\mathrm{d}}

\begin{document}

\title{Experimental Covert Communication Using Software-Defined Radio
\thanks{This work was supported, in part, by the National Science Foundation under Grants No.~CCF-2006679 and CNS-2107265.}
}

\author{\IEEEauthorblockN{Rohan Bali, Trevor E. Bailey, Michael S. Bullock, and Boulat A. Bash}
\IEEEauthorblockA{Department of Electrical and
Computer Engineering \\
University of Arizona, Tucson, AZ 85721, USA \\
\{rbali, trevorbailey, bullockm, boulat\}@arizona.edu}}

\maketitle

\begin{abstract}
The fundamental information-theoretic limits of covert, or low probability of detection (LPD), communication have been extensively studied for over a decade, resulting in the \emph{square root law} (SRL): only $L\sqrt{n}$ covert bits can be reliably transmitted over time-bandwidth product $n$, for constant $L>0$.
Transmitting more either results in detection or decoding errors.
The SRL imposes significant constraints on hardware realization of provably-secure covert communication.
Thus, experimental validation of covert communication is underexplored: to date, only two experimental studies of SRL-based covert communication are available, both focusing on optical channels. 
Here, we report our initial results demonstrating the provably-secure covert radio-frequency (RF) communication using software-defined radios (SDRs).
These validate theoretical predictions, open practical avenues for implementing covert communication systems, as well as raise future research questions.
\end{abstract}

\section{Introduction}

Warfighter operation in highly contested settings demands covert or low probability of detection/intercept (LPD/LPI) communication that enables  message transmission without alerting an adversary \cite{bash12sqrtlawisit, bash13squarerootjsacnonote, bash15covertcommmag, chen23covcommssurvey}. This contrasts the traditional cryptographic \cite{menezes96HAC} and information-theoretic secrecy \cite{bloch11pls} security that prevents access to the transmission's content, but not its detection.
Careful waveform design and spread spectrum techniques are often employed in practice to reduce adversaries' signal-to-noise ratio (SNR) below the noise floor \cite[Pt.~1, Ch.~5]{simon94ssh}.
However, guaranteeing covertness requires intricate mathematical analysis which yields the \emph{square root law} (SRL): only $B(n)=L\sqrt{n}$ covert bits can be reliably transmitted over $n$ channel uses \cite{bash12sqrtlawisit, bash13squarerootjsacnonote, bash15covertcommmag, chen23covcommssurvey}.
The channel-dependent constant $L>0$ is called \emph{covert capacity} and $n=TW$ is the transmission time-bandwidth product.
Notably, the associated Shannon capacity \cite{cover02IT} is zero, since $\lim_{n\to\infty}\frac{B(n)}{n}=0$.
This is because adversary in covert communication seeks just one bit of information (whether transmitter is on or not) versus $\mathcal{O}(n)$ bits of transmitted data in traditional secure communication.
Nevertheless, a significant number of such provably-covert bits can still be transmitted.

The discovery of the SRL in \cite{bash12sqrtlawisit, bash13squarerootjsacnonote} resulted in an explosion of research by the communication and information theory communities overviewed in a tutorial \cite{bash15covertcommmag} and a detailed survey \cite{chen23covcommssurvey}.
This includes characterization of capacity $L$ for additive white Gaussian noise (AWGN) and discrete memoryless channels (DMCs) \cite{bloch15covert, wang15covert, tahmasbi19covertdmc2ndorder, xinchun24secondorderawgncovert}, covert networks \cite{arumugam16broadcast,  hu17covertnetworks, arumugam18mac, tan18covertbc, azadeh18covertmultihop, soltani18netlpd}, quantum aspects of covert communication \cite{bash13quantumlpdisit, bash15covertbosoniccomm, azadeh16quantumcovert-isitarxiv, bullock20discretemod, gagatsos20codingcovcomm, tahmasbi19bosoniccovertqkd, tahmasbi20bosoniccovertqkd-jsait,
anderson21bosonicbroadast, wang22isitcoverttd, anderson2024covert-qce, bullock2025fundamentallimitscovertcommunication, zlotnick25eacovertcomm}, and many other directions.
Notwithstanding this progress on fundamental theory, experimental SRL-based covert communication remains underexplored, with only two published works, both focusing on optical channels \cite{bash15covertbosoniccomm, liu24metrofibercovcomm}.
This paper addresses this gap by providing the first, to our knowledge, experimental validation of the SRL on radio frequency (RF) channels.

We evaluate a covert communication protocol using software-defined radio (SDR).
Our implementation uses USRP X310 SDR units deployed on the ORBIT testbed \cite{raychaudhuri05orbit, orbit_grid}, as detailed in Section \ref{sec:Experiment_Setup}, enabling controlled and reproducible experiments.
Binary phase-shift keying (BPSK) with a Gaussian pulse shaping filter allows transmission with controlled temporal and spectral symbol leakage while mitigating timing jitter.
To meet the SRL, the transmitter uses a ``sparse coding'' strategy, selecting a random subset of available channel uses that is secretly shared with the receiver in advance.
An experimental framework based on a synthetic Gaussian noise source within a wired, shielded network provides environment control and supports reproducibility.

The SRL governs, arguably, the worst case scenario.
As detailed in Section \ref{sec:prerequisite}, it assumes that the characteristics of the transmitter-adversary channel (noise power variance and transmission loss), the exact time and frequency band of potential transmission, and details of the transceiver system design are known to the adversary.
However, the adversary 1) cannot control all the random channel noise; and, 2) lacks access to the secret shared between transmitter and receiver prior to the transmission.
We implement this model and show that one can transmit covertly under such conditions.
While conservative, it provides a high level of security to unforeseen adversarial technological surprises.
Indeed, relaxing these assumptions can result in significant performance gains.
For example, an adversary's uncertainty of transmission time/frequency yields a multiplicative improvement to covert capacity $L$ \cite{bash14timingisit, bash16timingtwc, arumugam16async}, while uncertainty in noise power level can lead to a linear law: $\mathcal{O}(n)$ covert bits reliably transmissible over $n$ channel uses \cite{sobers15jammer-asilomar, sobers17jammer}.
We defer validation of these to future work.
Additionally, our experiment motivates a systematic study of the impact of adversary hardware limitations, such as sampling timing jitter and receiver bandwidth constraints.
Furthermore, employing more efficient modulation and coding schemes (such as \cite{wang21covcodes}), as well as minimizing the pre-shared secret size (see \cite{bloch15covert}) are avenues for future exploration.

\begin{figure}[t!]
    \centering
    \includegraphics[width=1\linewidth]{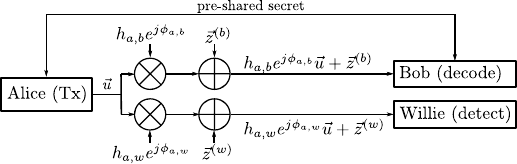}
    \caption[Covert communication over an AWGN channel.]{Covert communication over an AWGN channel. Alice ($a$) transmits a complex-valued covert message $\vec{u}\in\mathbb{C}^n$ over $n$ uses of a channel corrupted by independent AWGN at intended receiver Bob ($b$) and adversary (warden) Willie ($w$). Here $\vec{u}$ is a sequence of pulse-shaped symbols and empty pulse slots described in Section \ref{subsec:sysmodel}. Bob receives $h_{a,b}e^{j\theta_{a,b}}\vec{u} + \vec{z}^{(b)}$ while Willie observes $h_{a,w}e^{j\theta_{a,w}}\vec{u} + \vec{z}^{(w)}$. For $r\in \{b,w\}$, path loss $h_{a,r}$ and phase $\theta_{a,r}$ are static and known, while $\vec{z}^{(r)}$ is complex circularly-symmetric AWGN. Alice and Bob's pre-shared secret allows reliable message decoding while rendering it indistinguishable from noise $\vec{z}^{(w)}$ by Willie.}
    \label{fig:theoretical_channel}
\end{figure}

This paper is organized as follows: next we develop the theoretical foundation for RF covert communication by deriving a sparse-coded BPSK transmission scheme that satisfies the SRL for the discrete-time AWGN channel model. Section~\ref{sec:results} details the experiment on the ORBIT testbed, and presents our results. Finally, Section~\ref{sec:discussion} interprets these, highlighting practical design challenges and their implications for practical systems, and outlines future work. 
Appendices~\ref{ap:covertness}-\ref{ap:snr} provide supporting derivations.

\section{Theoretical System Analysis} \label{sec:prerequisite}

\subsection{Channel Model}
Consider a static discrete-time AWGN channel model depicted and described in Fig.~\ref{fig:theoretical_channel}. 
Per the \emph{square root law} (SRL), one can transmit $B(n)=L\sqrt{n}$ covert bits reliably in $n$ uses of such channel \cite{bash12sqrtlawisit, bash13squarerootjsacnonote}, with covert capacity $L$ derived in \cite{bloch15covert, wang15covert}. We adapt the results in \cite{bash12sqrtlawisit, bash13squarerootjsacnonote} to our SDR-based system model in the following subsections. 

\subsection{System Model and Reliable Covert Communication}\label{subsec:sysmodel}
The AWGN channel described in Fig.~\ref{fig:theoretical_channel} provides a discrete-time model of the covert communication scheme implemented in Section \ref{sec:Experiment_Setup}.
Analysis of the fundamental limits in such model \cite{bash12sqrtlawisit, bash13squarerootjsacnonote} often treats each channel use separately and assumes that Alice can use arbitrarily low output power.
However, practical radio systems require pulse-shaping for bandwidth efficiency, and to mitigate inter-symbol interference and timing jitter.
Thus, we divide $n$ available channel uses into $n_p\in\mathcal{O}(n)$ pulse slots, with pulse-shape vector $\vec{c}$ occupying $n_s=n/n_p>0$ channel uses.
Further, since minimum output energy is limited in practical radios, we fix the norm $\|\vec{c}\|$ and employ \emph{sparse coding} to ensure covertness: prior to transmission, Alice and Bob secretly share a random $n_p$-length sequence $\vec{t}$ of independent and identically distributed (i.i.d.) samples from the Bernoulli distribution: $p(t_i)=\{
1-\alpha_n \text{ if } t_i = 0; \alpha_n \text{ if } t_i = 1\}$. 
The number $n_t=\sum_{i=1}^{n_p} t_i$ of selected pulse slots is a random variable with average $\alpha_nn_p$. It is also the length of the transmitted message $\vec{x}_{n_t}$, in symbols.
BPSK modulates one bit per symbol, hence $\vec{x}_{n_t} \in \{-1,1\}^{n_t}$ is an $n_t$-bit vector.
Alice transmits either $\vec{c}$ or $-\vec{c}$ in pulse slot $i$ if $t_i$ is 1, and stays silent (transmitting $\vec{0}$, the innocent vector) otherwise. 
The probability of transmission  $\alpha_n \in \mathcal{O}\left(\frac{1}{\sqrt{n}}\right)$ follows the SRL \cite{bash12sqrtlawisit, bash13squarerootjsacnonote}.
We derive it in Section \ref{subsec:covertness}. We assume that  Alice and Bob have a reference that allows them to synchronize the start of transmission.

Alice transmits a symbol $x \in \{-1,1\}$ from $\vec{x}_{n_t}$ in  each of the $n_t$ selected pulse slots. Bob receives the $n_s$-sample vector $\vec{y}_p(x)=x h_{a,b} e^{j\theta_{a,b}}\vec{c} + \vec{z}^{(b)}$,
where $h_{a,b}$ and $\theta_{a,b}$ are the constant path loss and carrier-phase offset on Alice-to-Bob channel.
AWGN $\vec{z}^{(b)}$ is an i.i.d.~sample of complex circularly-symmetric Gaussian distribution $\mathcal{CN}\left(\vec{0},\sigma_b^2\mathbf{I}_{2n}\right)$, with $\mathbf{I}_{n}$ an $n\times n$ identity matrix.
We compensate for $\theta_{a,b}$ using single pilot symbol, per  Appendix~\ref{ap:phase_est}. Bob then estimates $\hat{x}$ from $\vec y_p(x)$ as $1$ if $\langle \vec{c}, \vec{y}_p(x)\rangle \geq \langle -\vec{c}, \vec{y}_p(x)\rangle$, and $-1$ otherwise.
Since AWGN is symmetric, this hard-decision scheme induces a binary symmetric channel (BSC) with probability of error: 
\begin{align}
     p_{e, \rm{bsc}}^{(b)}\triangleq \Pr(\hat{x}=-1|x=1)=\Pr(\hat{x}=1|x=-1).
\end{align}
For $n_p$ sufficiently large, Alice and Bob can use an error correction code (ECC) \cite{richardson2008moderncoding} on the $\approx\alpha_n n_p$-long subset of pulse slots $\{k:t_k=1\}$. 
This allows reliable transmission of
\begin{align}
    B_{\rm bsc}(n)&= n_tC_{\rm bsc} \approx \alpha_n n_p C_{\rm bsc}\label{eq:tp} 
\end{align}
bits in $n$ channel uses on average, with $C_{\rm bsc}\triangleq 1-h_2\left(p_{e, \rm{bsc}}^{(b)}\right)$, where $h_2(p)\triangleq-p\log_2(p)-(1-p)\log_2(1-p)$ is the binary entropy function and the approximation is due to $n_t$ being a random variable.

The ECC structure, which can aid adversary Willie\footnote{Although ``Eve'' is a typical adversary moniker in information security, here we use ``Warden Willie'' as is done in steganography \cite{fridrich09stego} to indicate the fundamentally different function of detection rather than eavesdropping.} in detection, is eliminated by applying an $n_t$-bit one-time-pad $\vec{s}$ to the encoder output.
Alice and Bob equiprobably select each bit in $\vec{s}$, resulting in output distribution $\Pr(x=-1)=\Pr(x=1)=\frac{1}{2}$. Pre-shared secret includes $\vec{t}$ and $\vec{s}$.

Alice and Bob share $\mathcal{O}(\sqrt{n}\log n)$ secret key bits, since $\vec{t}$ and $\vec{s}$ take $n_t\log_2 n_p$ and $n_t$ bits to represent, respectively. 
One can reduce the required number of secret bits to $\mathcal{O}(\sqrt{n})$ at a significant increase in complexity \cite{bloch15covert}.
However, computation using more energy than storage justifies this logarithmic cost for portable covert communication systems.

\subsection{Hypothesis Testing and Covertness} \label{subsec:covertness}
Willie has to decide whether Alice is transmitting based on observing $\vec{w}$. 
We assume that he knows the transmission start time, channel conditions, and other details of transceiver design (including $\vec{c}$ and $\alpha$). 
He cannot access $\vec{t}$ and $\vec{s}$.
He performs a binary hypothesis test between hypotheses $H_0$ (no transmission) and $H_1$ (transmission). 
Let distributions $P_0^{n}$ and $P_1^{n}$ and associated density functions $p_0^n(\vec{w})$ and $p_1^n(\vec{w})$ describe the statistics of Willie's output $\vec{w}$ when Alice is silent ($H_0$) and transmitting ($H_1$).
Assuming non-informative priors\footnote{Accounting for arbitrary priors is discussed in \cite{sobers17jammer}.} $\Pr(H_0)=\Pr(H_1)=\frac{1}{2}$, Willie's probability of error for an optimal detection scheme is \cite[Th.~13.1.1]{lehmann05stathyp}:
\begin{align}
    p_e^{(w)}=\frac{1}{2}-\frac{1}{2}\mathcal{V}_T(P_0^{n},P_1^n),\label{eq:tvbound}
\end{align}
where $\mathcal{V}_T(P_0^{n},P_1^n)\triangleq\frac{1}{2}\int_{\mathbb{R}^n}|p_0^n(\vec{w})-p_1^n(\vec{w})|\dif\vec{w}$ is the total variation distance between $P_0^{n}$ and $P_1^{n}$.
We say that the transmission is $\delta$-covert if $p_e^{(w)}\geq\frac{1}{2}-\delta$.
Total variation distance is mathematically unwieldy, so we employ Pinsker's inequality \cite[Lemma 11.6.1]{cover02IT} to lower bound 
\begin{align}
p_e^{(w)}
    &\geq \frac{1}{2}-\frac{1}{2\sqrt{2}}\sqrt{D(P_0^{ n}\|P_1^n)},\label{eq:pinskerbound}
\end{align}
where $D(P_0^{ n}\|P_1^n)\triangleq\int_{\mathbb{R}^n}p_0^n(\vec{w})\log_2\frac{p_0^n(\vec{w})}{p_1^n(\vec{w})}\dif\vec{w}$ is the relative entropy of $P_0^{n}$ and $P_1^{n}$.
Thus, instead of \eqref{eq:tvbound}, we use \eqref{eq:pinskerbound}, noting that any scheme is $\delta$-covert if $D(P_0^{ n},P_1^n)\leq\delta_{\rm RE}=8\delta^2$.

Alice inputs $x\in\{-1,0,1\}$ in each pulse slot (zero is silence).
Then, Willie receives $\vec{w}_p(x)=xh_{a,w}e^{\theta_{a,w}}\vec{c}+\vec{z}^{(w)}$, per the model in Fig.~\ref{fig:theoretical_channel}. Appendix~\ref{ap:covertness}  shows that setting
\begin{align}
    \alpha_n = \frac{2\sigma^2_w}{h_{a,w}^2\|\vec{c}\|^2}\sqrt{\frac{\delta_{\rm RE}}{n_p}}=2\mathrm{SNR}^{-1}\sqrt{\frac{\delta_{\rm RE}}{n_p}},\label{eq:anreq}
\end{align}
where $\mathrm{SNR}\triangleq \frac{h_{a,w}^2\|\vec{c}\|^2}{\sigma^2_w}$ is Willie's received SNR, ensures $\delta$-covertness for $\delta_{\rm RE}=8\delta^2$ of the transmission scheme described in Section \ref{subsec:sysmodel}. 
Combining 
\eqref{eq:anreq} with
$n_p=n/n_s$ and \eqref{eq:tp} yields the SRL scaling $B_{\rm bsc}(n)=\frac{2 C_{\rm bsc}\sqrt{\delta_{\rm RE}n}}{\mathrm{SNR}\times \sqrt{n_s}}\in\mathcal{O}(\sqrt{n})$. 
We note that the efficiency of covert communications over SDRs can be improved, as $\frac{2 C_{\rm bsc}\sqrt{\delta_{\rm RE}}}{\mathrm{SNR}\times \sqrt{n_s}}$ is significantly smaller than covert capacity $L$ of AWGN channel.
We discuss future work addressing this in Section \ref{sec:discussion}.

\section{Experiment Implementation and Results}
\label{sec:results}
\subsection{System Configuration}
\label{sec:Experiment_Setup}

\begin{figure}
    \centering
    \includegraphics[width=1\linewidth]{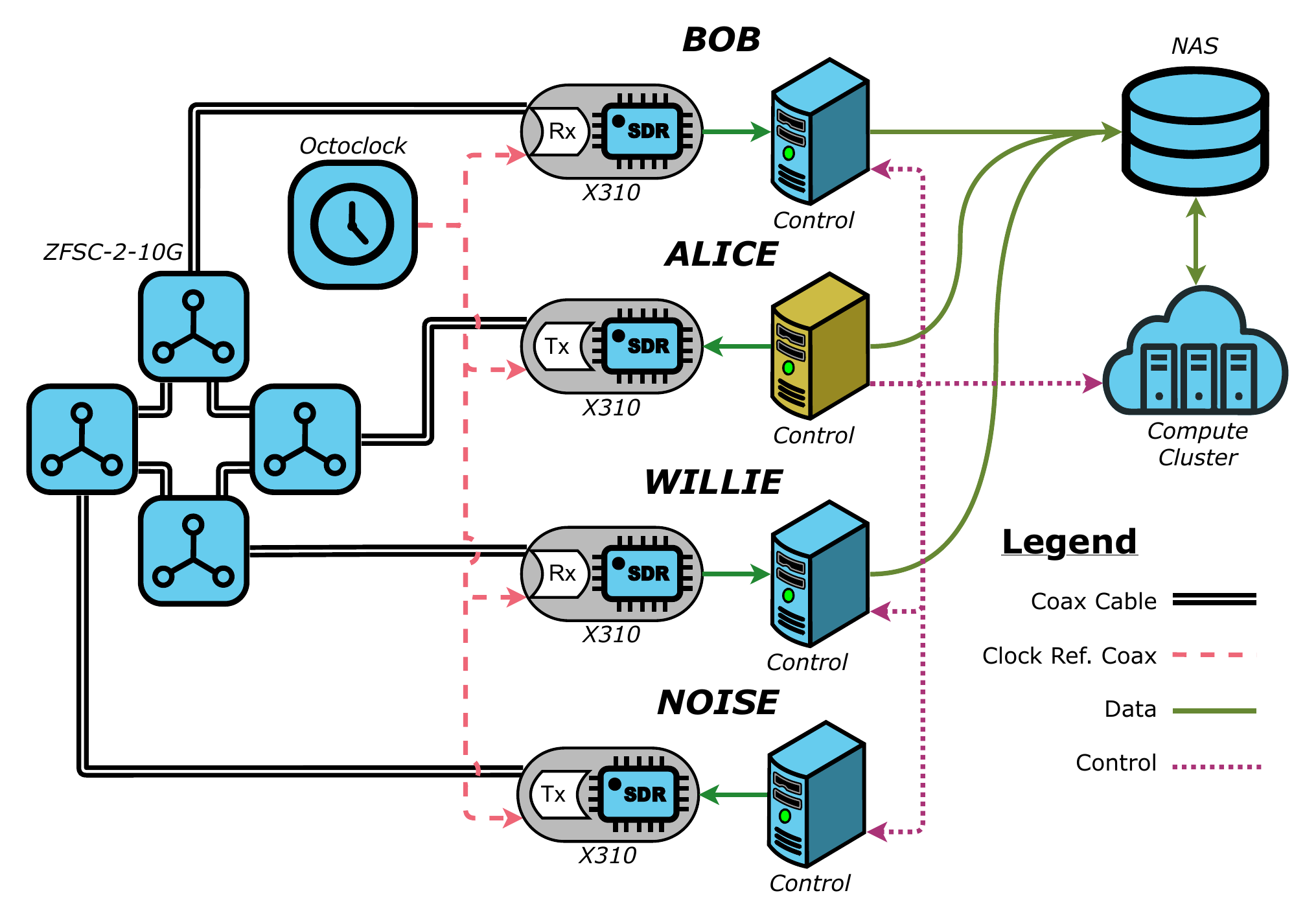}
    \caption{Covert communication experiment on ORBIT. Four Ettus USRP X310 radios -- Alice (Tx), Bob (Rx), Willie (warden), and a broadband noise source -- are linked by coaxial cables in a star topology using Mini-Circuits ZFSC-2-10G splitters/combiners. Tx-to-Rx and Tx-to-Tx path losses are 50 dB and 65 dB, respectively. All radios operate at \(f_c = 915\,\text{MHz}\) with Alice, Bob, and Willie using a DAC/ADC sampling rate of \(f_s = 12.5\,\text{mega-samples/s}\)  and the noise generator $f_n = 25\, \text{mega-samples/s}$. Alice, Bob, and Willie apply 0 dB Tx/Rx gain while the noise generator applies 20 dB gain. Each X310 connects over a 10 Gb/s SFP+ link to a dedicated control node (Intel Xeon E5-2640, 20 cores); Alice’s node orchestrates the experiment via TCP messages to the other nodes and an eleven-node compute cluster of the same machines that performs real-time processing while a 2 TB network attached storage (NAS), mounted via NFS v4.2, provides shared buffer.}
    \label{fig:detailed_schematic}
\end{figure}

We use ORBIT, an open-access radio grid testbed \cite{raychaudhuri05orbit, orbit_grid}.
As depicted in Fig.~\ref{fig:detailed_schematic}, we connect four Ettus universal software radio peripheral (USRP) X310 SDRs fitted with UBX daughterboards -- corresponding to Alice,  Bob,  Willie, and an AWGN source -- into a shared-medium RF star network topology via coaxial cables.
This isolates our experiment from others on ORBIT, enables control over the environment, and allows repeatability.
We defer a wireless experiment with an organic noise source to future work.
The radios are placed so that Bob and Willie experience approximately equal attenuation from both the AWGN generator and Alice, corresponding to an adverse scenario of Willie being close to Bob.
We use $f_s=12.5$ mega-sample/s digital-to-analog/analog-to-digital converter (DAC/ADC) sampling rate at Alice, Bob, and Willie's radios, and $f_n=25$ mega-sample/s at noise generator. We utilize a single channel centered at $f_c=915$ MHz of bandwidth no more than $W=6.25$ MHz.

Each radio has a dedicated enhanced small form-factor pluggable (SFP+) cable connecting it to a high-bandwidth router, and the router to a dedicated control node on the ORBIT grid \cite{orbit_grid}.
SFP+ ports operate in 10 Gbps mode to support the maximum transmission unit (MTU) size of 8 kB.
This prevents packet drops from the internal radio buffer overflows.
We note that even a single packet drop causes catastrophic misalignment of the covert symbols within the transmission, rendering useless an entire experimental trial. 
ORBIT provides a network \cite{orbit_grid} connecting the control nodes to network attached storage (NAS) and to other nodes forming a compute cluster for processing the collected data.

The radios' internal clocks are synchronized using an Ettus OctoClock, which provides low-jitter pulse-per-second (PPS) and 10 MHz reference signals, allowing the radios to maintain a constant phase offset \cite{octoclock_datasheet}. 
We note that a centralized clock source is merely an experimental convenience.
In practice, Alice and Bob can synchronize their own \emph{independent} stable time sources, such as atomic clocks, prior to transmission.\footnote{In a separate experiment in our lab at the University of Arizona, we confirmed this using a Stanford Research Systems FS725 10 MHz Rubidium Frequency Standard connected to Ettus USRP N210 radios.}

\begin{figure*}[!t]
\centering
\includegraphics[width=0.95\textwidth]{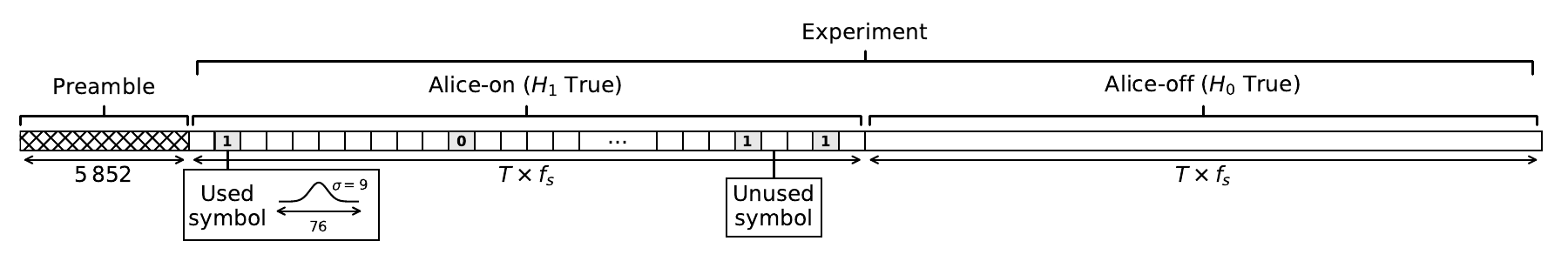}
\caption{ 
Data-packet structure (durations in samples). BPSK modulates one bit per symbol using 76 samples: a sample modulating the bit followed by the 75-sample zero pad. Preamble uses five 13-bit Barker sequences (65 bits total) and is pulse-shaped with a 913-sample RRC filter, yielding a 5\,852-sample header. The subsequent $T$-second, $T\times f_s$-sample, segments encode the two parts of the experimental trial. In the Alice-on segment, Alice transmits 76-sample Gaussian pulse-shaped symbols encoding random bits in pulse slots (indicated by shading) randomly chosen and stored in $\vec{t}$ (see Section \ref{subsec:sysmodel}); in the second (Alice-off) segment, she remains silent.}
\label{fig:data_packet}
\end{figure*}

Alice’s node generates fresh pre-shared secret and message vectors $\vec{t}$ and $\vec{x}_{n_t}$ for each experimental trial. Bob’s and Willie’s radios sample the channel continuously while Alice brackets the trial packet described in Section \ref{subsec:radio_comms} with TCP control messages that mark its precise start and stop times. At the end of each trial, control nodes write their data to the network-attached storage (NAS): Alice's node stores $\vec{t}$ and $\vec{x}_{n_t}$, whereas Bob’s and Willie’s nodes write their raw samples, padded with brief pre- and post-buffers. Alice's node then alerts the compute cluster, which analyzes the new trial, and deletes its data on completion. Acting as a rolling buffer, the NAS keeps heavy data traffic off the radio-control links and holds disk usage well below its 2 TB capacity even though the experiment produces 4.9 TB overall. The Ettus USRP library and USRP hardware driver (UHD), \cite{uhd_github, uhd_manual} are used to interact with the radios.
\subsection{Transmission Structure for Experimental Trials} 
\label{subsec:radio_comms}

In each experimental trial, Alice transmits a three-segment packet shown in Fig.~\ref{fig:data_packet}: a non-covert \emph{preamble}, an \emph{Alice-on} segment, and an \emph{Alice-off} (noise-only) segment. The last two segments are each $T$ seconds long. The preamble is the 13-bit Barker code repeated five times (65 bits total) for indicating the beginning of each experimental trial (which Bob and Willie locate by match-filtering) and synchronizing time.
The Alice-on segment carries Alice’s hidden message.
The AWGN noise generator is active for the entire experiment.

We employ BPSK, modulating one bit per symbol.
Each symbol has 76 samples: first sample modulating the bit and 75 zero-pad samples.
We apply a 12-tap root-raised-cosine (RRC) pulse-shaping filter (represented digitally by $12\times 76+1=913$ samples) with roll-off factor $\beta = 0.35$ to the preamble, using $65\times76+(913-1)=5\,852$ samples.

We evaluate the number of covert bits that can be reliably received by Bob and Willie's detector performance when hypothesis $H_1$ is true using the \emph{Alice-on} segment.  
Alice uses the pulse slots indicated in $\vec{t}$, and a 37-sample Gaussian pulse-shaping filter with $\sigma = 9$ samples. 
Since 99.5\% of a pulse's energy is contained within the filter's 37 samples, the energy of the generated pulse is contained within the 76-sample pulse slot. 
The \emph{Alice-off}  segment enables evaluation of Willie's detector's performance when hypothesis  $H_0$ is true.

\subsection{Experiment design}
\label{subsec:exp_design}

We select ten logarithmically-spaced values of Alice's transmission duration $T\in[0.05,15]$ s.
For each $T$, we conduct $N=500$ independent trials using distinct transmission packets described in Section \ref{subsec:radio_comms}.
Before we begin our experiment, we estimate Willie’s SNR using a single calibration transmission with a modified version of the data-packet described in subsection~\ref{subsec:radio_comms}: every fifth pulse slot in the Alice-on segment is used to transmit a random bit and Alice-off segment is deleted.
We outline our estimator for SNR in Appendix~\ref{ap:snr}.
We use the SNR estimate to compute $\alpha_n\sqrt{n}=\frac{4\sqrt{2}\delta}{\mathrm{SNR}}$.
We set $\delta=0.05$ and use the result to compute $\alpha_n$ for each value of $T$.
This is used to generate random vectors $\vec{t}$ with pulse locations used by Alice to transmit for each trial.

\subsection{Results}
\label{subsec:results}

\begin{figure}[!th]
\centering
\includegraphics[width=1\columnwidth]{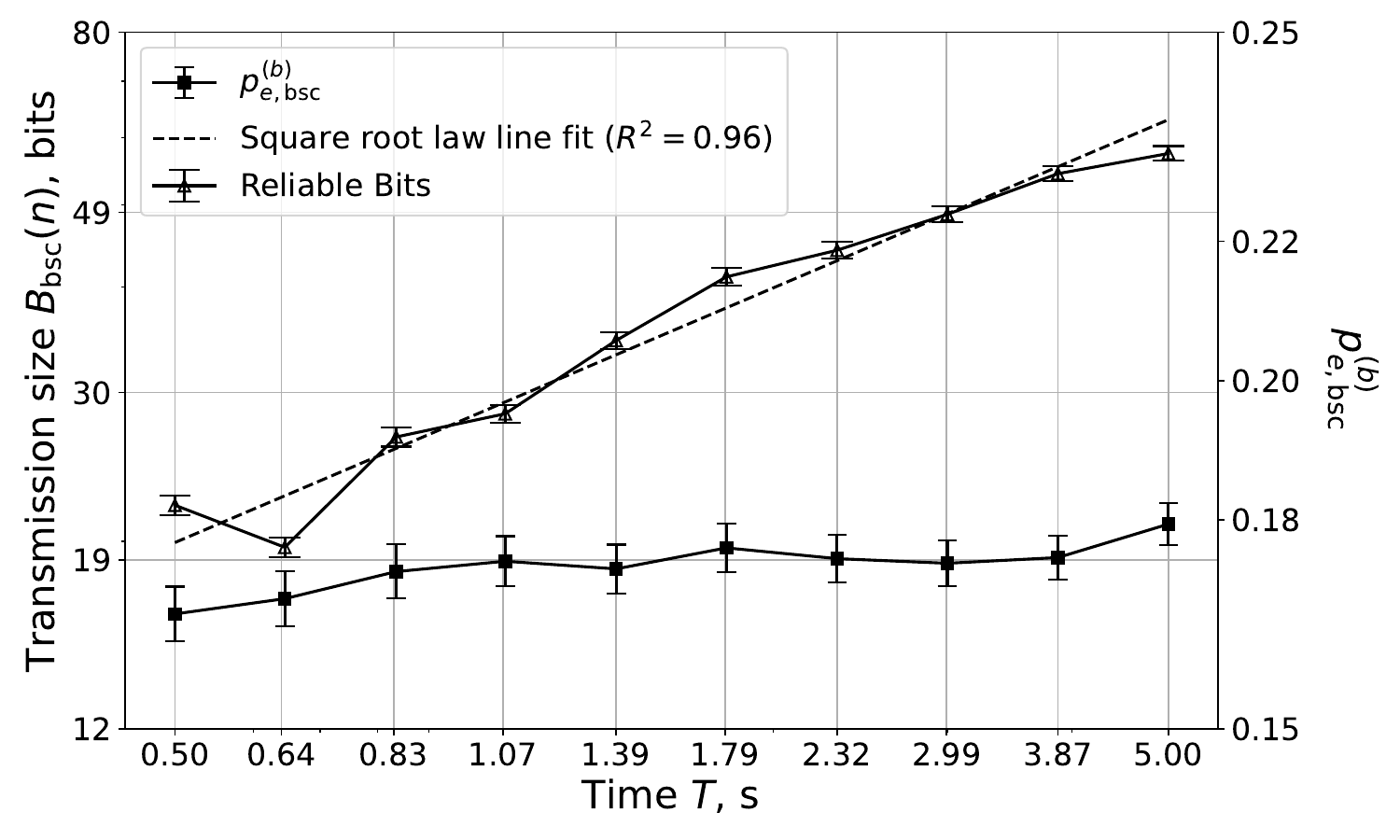}
\caption{Bob's receiver performance. Total number $B_{\mathrm{bsc}}(n)$ of reliably-decodable bits (left ordinate) and decoding error probability $p_{e,\mathrm{bsc}}^{(b)}$ (right ordinate) are plotted vs.~transmission duration $T$, with 95\% confidence intervals as error bars given $N=500$ trials per datapoint. Abscissa and left ordinate use a logarithmic scale; right ordinate uses a linear scale. Note that the left ordinate ranges from 12 to 80, and the right ranges from 0.15 to 0.25. Number of channel uses is $n=f_s T$. }
\label{fig:bob_bits_n_pe}
\end{figure}

Fig.~\ref{fig:bob_bits_n_pe} plots Bob's receiver's performance vs.~transmission duration $T$.
Bob uses $\vec{t}$ to estimate bits only in the pulse locations that Alice uses for transmission, per sparse coding described in Section \ref{subsec:sysmodel}.
We report the decoding error probability $p_{e,\mathrm{bsc}}^{(b)}$ estimated by averaging over $N=500$ trials using the right ordinate of Fig.~\ref{fig:bob_bits_n_pe}.
We observe that $p_{e,\mathrm{bsc}}^{(b)}\approx 0.17$ throughout our experiments.
Using the left ordinate we report the corresponding estimate of the total number of transmissible covert bits $B_{\rm bsc}(n)$ using the equality in \eqref{eq:tp}.
Fitting a line with slope of one-half to the log-log plot results in the coefficient of determination $R^2=0.96$, indicating the SRL-scaling that we expect.

\begin{figure}[!th]
\centering
\includegraphics[width=1\columnwidth]{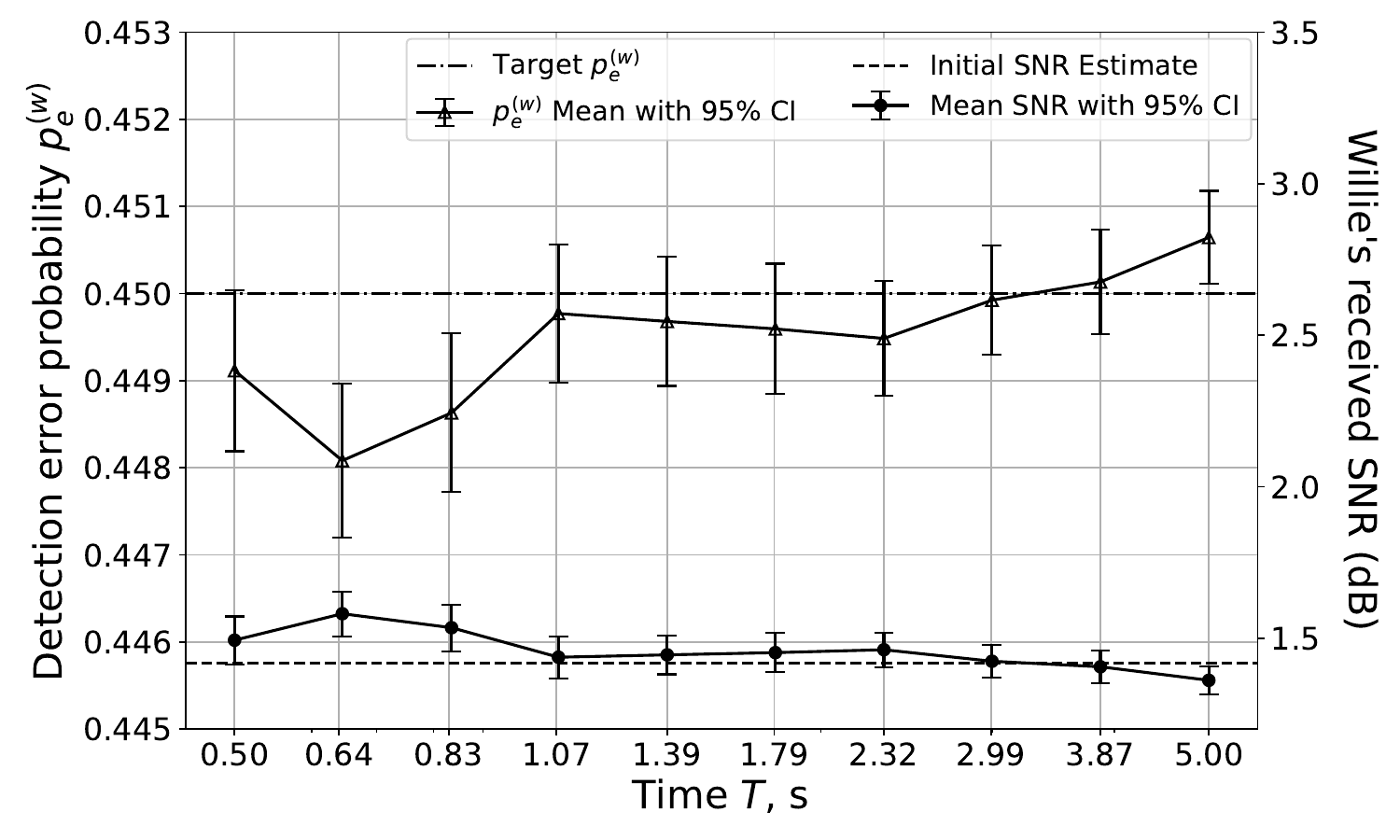}
\caption{Willie's detector performance. Detection probability error $p_e^{(w)}$ (left ordinate) and SNR in dB (right ordinate) are plotted vs.~transmission duration $T$, with 95\% confidence intervals shown as error bars given $N=500$ trials per datapoint. Note that the left ordinate ranges from 0.445 to 0.453, and the right ranges from 1.2 to 3.5. The abscissa uses a logarithmic scale. }
\label{fig:willie_snr_pe}
\end{figure}

Fig.~\ref{fig:willie_snr_pe} presents Willie's detector performance.
We employ the estimator from Appendix~\ref{ap:snr} to estimate Willie's received SNR, which we plot using the right ordinate.
We note that it remains close to the initial estimate through the duration of the experiment.
We then estimate the lower bound on Willie's probability of error $p_e^{(w)}$ by computing the upper bound on relative entropy derived from Taylor series expansion in  Appendix~\ref{ap:covertness} (and verifying that it is indeed an upper bound per Remark \ref{rem:taylor} therein).
We plot it using the left ordinate and note that it is very conservative, as it is, effectively, a lower bound on a lower bound.
Nevertheless, this is sufficient to show that we indeed achieve covert communication.
Next, we discuss follow-on work that includes further investigation of Willie's receiver.

\section{Conclusion, Discussion, and Future Work} \label{sec:discussion}

We demonstrate the first implementation of SRL-based covert communication in RF domain.
We employ SDRs, which are not specifically designed for covert communication. We have to address significant challenges:
\begin{itemize}
    \item \emph{Dynamic range and ADC/DAC granularity limitations}: The UBX daughterboards use a 16-bit DAC for transmission and a 14-bit ADC for reception. While Alice can generate high-resolution low-power pulses, Bob's lower ADC precision limits his ability to distinguish weak signals from noise. Ensuring detection of pulses after digitization leads to minimum transmission power, requiring the use of sparse coding. This, in turn, needs tight time synchronization, which we demonstrate.
    \item \emph{Time and frequency synchronization}: The oven-controlled crystal oscillators (OCXOs) in USRP X310s lack the frequency accuracy and stability needed to decode sparsely transmitted covert symbols. However, as the covert symbol pattern is sparse, conventional methods, such as Costas loops, do not converge reliably. Furthermore, Bob has to know precisely when the Alice begins transmitting. We utilize a common reference (OctoClock) and a non-covert preamble, however, in practice, GPS or a highly stable time source such as atomic clock can be used for both disciplining the local oscillators and timing information. We will employ these in follow-on experiments. 
    \item \emph{Phase synchronization and channel state information (CSI)}: Here, Alice corrects the global phase offset using a single pilot symbol. In practice, mobility requires more frequent phase correction as well as estimation of CSI. Non-covert communication systems can transmit periodic pilot symbols. SRL renders this ineffective in covert systems, however, blind methods \cite[Sec.~IV]{tsatsanis95tvsystems} may be adapted. We defer this to future work.
    \item \emph{Data volume:} We capture $\approx 7$\,h of baseband data at $f_s = 12.5$\,MHz with 64-bit in-phase and quadrature samples for each receiver (Bob and Willie), yielding $\approx 4.9$ TB on disk. While we employ real-time processing on the compute cluster, the computational and storage burden must be reduced for practical systems.
    \item \emph{Continuous noise injection}: Maintaining an always-on, spectrally flat artificial noise floor is limited by the maximum DAC sampling rate the system can sustain: sampling too fast can overwhelm the host CPU and lead to buffer under-flows. In the future, we will explore using more than one radio to emulate noise as well as more natural noise sources by experimenting ``in the wild.''
\end{itemize}

Indeed, our proof-of-concept experiment raises many theoretical and experimental research questions.
In the short term, we plan to address some of them by improving our ORBIT-based design as follows:
\begin{itemize}
    \item We will increase our communication system efficiency by optimizing the length $n_s$ of our pulse-shape vector $\vec{c}$ as well as employing quadrature phase-shift keying (QPSK) instead of BPSK;
    \item We will add control over path loss and phase shift to introduce channel dynamics;
    \item We will estimate Willie's detection error probability directly by estimating the output distributions for the optimal test statistics, as is done in \cite{bash15covertbosoniccomm}.
\end{itemize}
Additionally, AWGN channel model provides only a zeroth-order approximation to practical RF channels.
Thus, SRL-based covert communication needs to be validated in a more realistic, dynamic environment.
We plan on evolving our system to be independent rather than centrally-controlled.
This would allow not only experimentation ``in the wild'' but also exploration of covert networks.
Furthermore, we plan on relaxing assumption on adversary's capabilities and studying the impact of, e.g., lack of precise knowledge of the timing of the transmission and pulse shape used.

\section*{Acknowledgment}
The authors are grateful to Ivan Seskar and ORBIT staff for setting up the radios on ORBIT.
The authors thank Loukas Lazos, Ming Li, Jingcheng Li, Ziqi Xu, Samuel H.~Knarr, and Timothy C.~Burt for valuable advice, and sharing the equipment.
Finally, the authors acknowledge helpful discussions with Mark J.~Meisner, Jaim Bucay, Dennis L.~Goeckel, Donald F.~Towsley, Matthieu R.~Bloch, Matthew Arcarese, and Robert J.~McGurrin.

\appendices

\section{Covertness Criterion Analysis}\label{ap:covertness}
Since Willie knows the start time, the duration of Alice's transmission, and details of her system from Section \ref{subsec:sysmodel}, he collects $n$ observations corresponding to the total number of channel uses available to Alice.
Furthermore, we assume phase $\theta_{a,w}=0$, and allow Willie to discard the quadrature components of his observations, which contain only noise, leaving him with with in-phase components which may have Alice's BPSK-modulated symbols.
These $n$ observations are divided into $n_p$ segments of $n_s$ observations each, corresponding to pulse slots.
Each segment $\vec{w}^{(h)}_i$ is indexed according to the true hypothesis $H_h$, $h\in\{0,1\}$ and pulse location $i=1,\ldots,n_p$. Denote by $\phi(\vec{x};\vec{\mu}, \mathbf{\Sigma})$ the multi-dimensional Gaussian density function with mean vector $\vec{\mu}$ and covariance matrix $\mathbf{\Sigma}$.
Under $H_0$, Alice does not transmit and Willie observes AWGN. Thus, segments are i.i.d.~with the density function:
\begin{align}
    p\left(\vec{w}^{(0)}_i\right)&=\phi\left(\vec{w}^{(0)}_i;\vec{0}, \sigma^2_w \mathbf{I}_{n_s}\right).\label{eq:prbpulsewillie0}
\end{align}
Alice transmits under hypothesis $H_1$.
Since Willie does not have $\vec{t}$ and $\vec{s}$, segments are i.i.d.~with the density function:
\begin{align}
    p\left(\vec{w}^{(1)}_i\right)\nonumber&=(1-\alpha_n)\phi\left(\vec{w}^{(1)}_i;\vec{0}, \sigma^2_w \mathbf{I}_{n_s}\right) \\&\phantom{=}+ \frac{\alpha_n}{2}\phi\left(\vec{w}^{(1)}_i;h_{a,w}\vec{c}, \sigma^2_w \nonumber\mathbf{I}_{n_s}\right)\\&\phantom{=}+ \frac{\alpha_n}{2}\phi\left(\vec{w}^{(1)}_i;-h_{a,w}\vec{c}, \sigma^2_w \mathbf{I}_{n_s}\right).\label{eq:prbpulsewillie}
\end{align}
Note that additivity of relative entropy implies $D(P_0^{n}\|P_1^n)=n_p D(P_0^{n_s}\|P_1^{n_s})$, where \eqref{eq:prbpulsewillie0} and \eqref{eq:prbpulsewillie} are the respective density functions for distributions $P_0^{n_s}$ and $P_1^{n_s}$.
Similar to \cite[Th.~1.2]{bash13squarerootjsac}, we take the Taylor series expansion of $D(P_0^{n_s}\|P_1^{n_s})$ at $\|c\|= 0$. The first three terms are zero. For the fourth term we need:
\begin{align}
    &\frac{{\dif}^4D(P_0^{n_s}\|P_1^{n_s})}{{\dif}\|c\|^4} \nonumber \\&\phantom{=}=\nonumber \frac{\dif}{{\dif} \|c\|^4}\int  \dif\vec{w}_{i}^{(1)}\frac{e^{-\frac{1}{2\sigma^2_w}\|\vec{w}_{i}^{(0)}\|^2}}{(2\pi \sigma^2_w)^{n_s/2}}\log\left(1-\alpha_n\vphantom{\exp\left(-\frac{h_{a,w}^2\|\vec{c}\|^2}{2\sigma_w^2}\right)}\right.\\&\phantom{=====
    ===}\left.+\alpha_n e^{\left(-\frac{h_{a,w}^2\|\vec{c}\|^2}{2\sigma_w^2}\right)}\cosh\left(h_{a,w}\left\langle \vec{w}_{i}^{(1)}\middle |\vec{c}\right\rangle\right)\right).
\end{align}
Adapting the argument in \cite[App.~A]{bash13squarerootjsac} yields:
\begin{align}
    \left.\frac{\|c\|^4}{4!}\left(\frac{{\dif }^4D(P_0^{n_s}\|P_1^{n_s})}{{\dif}\|c\|^4}\right\lvert_{\|c\|=0}\right)
    =\frac{\alpha_n^2  h_{a,w}^4\|c\|^4}{4\sigma_w^4}\label{eq:relent4th}.
\end{align}
Using Taylor's theorem with remainder and rearranging terms yields \eqref{eq:anreq}.
\begin{remark}
\label{rem:taylor}
    The argument using Taylor's theorem with remainder from \cite[Th.~1]{bash13squarerootjsac} is contingent on the BPSK signal power being arbitrarily small. We may adapt this argument for our experimental setup by showing that the sixth term in the expansion is negative for every $\xi\in\left[0,\|\vec{c}\|\right]$. We verify this numerically for our estimated experimental parameters but omit the details for brevity.
\end{remark}

\section{Phase Estimation}\label{ap:phase_est}

A pulse-bearing slot corresponding to a pilot symbol received by Bob is:
\begin{align}
\vec{y}_p(x)&=xh_{a,b}e^{j\theta_{a,b}}\vec{c}+\vec{z}^{(b)},
\end{align}
for known $x\in\{-1,1\}$ and unknown phase $\theta_{a,b}$.
Applying the pulse-shape filter and multiplying by $x$ yields:
\begin{align}
x\langle \vec{c},\vec{y}_p(x)\rangle&=h_{a,b}e^{j\theta_{a,b}}\|\vec{c}\|^2+x\langle \vec{c},\vec{z}^{(b)}\rangle.
\end{align}
The expected values of the in-phase and quadrature (IQ) components $p_I\triangleq \Re\left(x\langle \vec{c},\vec{y}_p(x)\rangle\right)$ and $p_Q\triangleq \Im\left(x\langle \vec{c},\vec{y}_p(x)\rangle\right)$ of $x\langle \vec{c},\vec{y}_p(x)\rangle$ are:
\begin{align}
    \mathbb{E}\left[\Re\left(x\langle \vec{c},\vec{y}_p(x)\rangle\right)\right] &=\mathbb{E}[p_I] = h_{a,b}\|\vec{c}\|^2\cos(\theta_{a,b})\\
    \mathbb{E}\left[\Im\left(x\langle \vec{c},\vec{y}_p(x)\rangle\right)\right] &=\mathbb{E}[p_Q] = h_{a,b}\|\vec{c}\|^2\sin(\theta_{a,b}),
\end{align}
since AWGN is circularly-symmetric and has zero mean.
Thus, averaging over many instances of $p_I$ and $p_Q$ yields their estimates $\hat{p}_I$ and $\hat{p}_Q$.
The estimate of phase $\theta$ is then $\hat{\theta}=\tan^{-1}\frac{\hat{p}_I}{\hat{p}_Q}$.
We note that, while in practical communication systems $\theta$ evolves, and requires many pilot symbols (or blind methods \cite[Sec.~IV]{tsatsanis95tvsystems}) to track, here just one pilot symbol is sufficient to accurately estimate it.
We defer investigation of mitigating the impact of phase dynamics in covert communication to future work.

\section{Estimation of Willie's SNR}\label{ap:snr}

We need to estimate Willie's SNR given the knowledge of $\vec{c}$, $n_s$, as well as transmitted symbol in $\vec{x}_{n_t}$ and their locations $\vec{t}$.
While Willie has no access to $\vec{x}_{n_t}$ and $\vec{t}$, we use them to characterize the SNR of his system.
A pulse-bearing slot received by Willie is:
\begin{align}
\vec{w}_p(x)&=xh_{a,w}e^{j\theta_{a,w}}\vec{c}+\vec{z}^{(w)},
\end{align}
for $x\in\{-1,1\}$ and unknown phase $\theta_{a,w}$.
Applying the pulse-shape filter and multiplying by $x$ yields:
\begin{align}
x\langle \vec{c},\vec{w}_p(x)\rangle&=h_{a,w}e^{j\theta_{a,w}}\|\vec{c}\|^2+x\langle \vec{c},\vec{z}^{(w)}\rangle.
\end{align}
The expected value of the above is:
\begin{align}
    \mathbb{E}\left[x\langle \vec{c},\vec{w}_p(x)\rangle\right]&=h_{a,w}e^{j\theta_{a,w}}\|\vec{c}\|^2,
\end{align}
since AWGN has zero mean, and, hence, $\mathbb{E}\left[x\langle \vec{c},\vec{z}^{(w)}\rangle\right]=0$.
Thus, averaging over many instances of pulse-bearing slot observations, and dividing by a known constant $\|\vec{c}\|^2$, yields an estimate of $h_{a,w}e^{j\theta_{a,w}}$.
Squared magnitude of this is the estimate $\hat{h}_{a,w}^2$ of $h_{a,w}^2$.

An empty pulse slot received by Willie contains only noise:
\begin{align}
\vec{w}_p(0)&=\vec{z}^{(w)},
\end{align}
The expectation of its squared magnitude is:
\begin{align}
    \mathbb{E}\left[\langle \vec{w}_p(0),\vec{w}_p(0)\rangle\right]&=\mathbb{E}\left[\langle \vec{z}^{(w)},\vec{z}^{(w)}\rangle\right]=n_s\sigma_w^2.
\end{align}
Thus, averaging over many instances of empty pulse slots, and dividing by a known constant $n_s$, yields an estimate $\hat{\sigma}_w^2$ of $\sigma_w^2$.
Finally, we estimate the SNR by evaluating $\frac{\hat{h}_{a,w}^2\|c\|^2}{\hat{\sigma}_w^2}$.


\end{document}